# Case of two electrostatics problems: Can providing a diagram adversely impact introductory physics students' problem solving performance?

Alexandru Maries[1] and Chandralekha Singh[2]

[1]*Department of Physics, University of Cincinnati, Cincinnati, Ohio 45221, USA*
[2]*Department of Physics and Astronomy, University of Pittsburgh, Pittsburgh, Pennsylvania 15260, USA*

Drawing appropriate diagrams is a useful problem solving heuristic that can transform a problem into a representation that is easier to exploit for solving it. One major focus while helping introductory physics students learn effective problem solving is to help them understand that drawing diagrams can facilitate problem solution. We conducted an investigation in which two different interventions were implemented during recitation quizzes in a large enrollment algebra-based introductory physics course. Students were either (i) asked to solve problems in which the diagrams were drawn for them or (ii) explicitly told to draw a diagram. A comparison group was not given any instruction regarding diagrams. We developed rubrics to score the problem solving performance of students in different intervention groups and investigated ten problems. We found that students who were provided diagrams never performed better and actually performed worse than the other students on three problems, one involving standing sound waves in a tube (discussed elsewhere) and two problems in electricity which we focus on here. These two problems were the only problems in electricity that involved considerations of initial and final conditions, which may partly account for why students provided with diagrams performed significantly worse than students who were not provided with diagrams. In order to explore potential reasons for this finding, we conducted interviews with students and found that some students provided with diagrams may have spent less time on the conceptual analysis and planning stage of the problem solving process. In particular, those provided with the diagram were more likely to jump into the implementation stage of problem solving early without fully analyzing and understanding the problem, which can increase the likelihood of mistakes in solutions.

## I. INTRODUCTION

Physics is a challenging subject to learn and it is especially difficult for introductory students to associate the abstract concepts they study in physics with more concrete representations that facilitate understanding without an explicit instructional strategy aimed to aid them in this regard. Here, by "representation," we mean any of the diverse forms in which scientific knowledge, or, physical concepts to be more exact, are understood and communicated [1]. This very broad definition encompasses nearly anything scientists have used to describe the world, but to be a bit more precise, in this article, we specifically refer to verbal, diagrammatic, mathematical, and graphical representations (for more information on how these are defined see Ref. [1]). Without guidance, introductory students often employ formula oriented problem solving strategies instead of developing a solid grasp of physical principles and concepts [2–5]. There are many reasons why multiple representations of concepts along with the ability to construct, interpret, and transform between different representations that correspond to the same physical system or process play a positive role in learning physics. First, physics experts often use multiple representations as a first step in a problem solving process [2,6–11]. Second, students who are taught explicit problem solving strategies emphasizing use of different representations of knowledge at various stages of problem solving construct higher quality and more complete representations and perform better than students who learn traditional problem solving strategies [12,13]. Third, multiple representations are very useful in translating the initial, usually verbal description of a problem into a representation more suitable to further analysis and mathematical manipulation [6,14–18] partly because the process of constructing an effective representation of a problem makes it easier to generate appropriate decisions about the solution process. Also, getting students to represent a problem in different ways helps shift their focus from merely manipulating equations toward understanding physics [19–22]. Some researchers have argued that in order to understand a physical concept thoroughly, one needs to be able to recognize and manipulate the concept in a variety of representations [14,23]. As Meltzer puts it [1], a range of diverse representations is required to "span" the conceptual space associated with an idea. Since traditional courses which do not emphasize multiple representations lead to low gains on the Force Concept Inventory [24,25] and on other assessments in the domain of electricity and magnetism [26–28], in order to improve students' understanding of physics concepts, many researchers have developed instructional strategies that place explicit [8,14,29–32] or implicit [16,33–40] emphasis on multiple representations. Van Heuvelen's approach, for example [14], starts by ensuring that students explore the qualitative nature of concepts by using a variety of representations of a concept in a familiar setting before adding the complexities of mathematics. Many other researchers have emphasized the importance of students becoming facile in translating between different representations of knowledge [29,41–47] and that significant positive learning occurs when students develop facility in the use of multiple forms of representation [48–51]. However, careful attention must be paid to instructional use of diverse representational modes since specific learning difficulties may arise as a consequence [1] because students can approach the same problem posed in different representations differently without support [1,49,52,53].

One representation useful in the initial conceptual analysis and planning stages of a solution is a schematic diagram of the physical situation presented in the problem (here, we mean a diagram used to visualize the problem which does not need to include any physics-specific details). Diagrammatic representations have been shown to be superior to exclusively employing verbal representations when solving problems [9,10,17,18]. It is therefore not surprising that physics experts automatically employ diagrams in attempting to solve problems [6,23,54,55]. However, introductory physics students need explicit help to (i) understand that drawing a diagram is an important step in organizing and simplifying the given information into a representation which is more suitable to further analysis [56], and (ii) learn to draw appropriate and useful diagrams. Therefore, many researchers who have developed strategies for teaching students effective problem solving skills use scaffolding support designed to help students recognize how important the step of drawing a diagram is in solving physics problems and guidance to help them draw useful diagrams. In Newtonian mechanics, Reif [2,8] has suggested that several diagrams be drawn: one diagram of the problem description, which includes all objects, and one diagram for each system that needs to be considered separately. Also, he described in detail concrete steps that students should take in order to draw these diagrams as follows:
 (a) describe both motion and interactions,
 (b) identify interacting objects before identifying forces,
 (c) separate long range and contact interactions, and
 (d) label contact points by the magnitude of the action-reaction pair of forces.

Van Heuvelen's active learning problem sheets (ALPS) [14] adapted from Reif follow a very similar underlying approach. Other researchers who have emphasized, among other things, the importance of diagrams in their approach to teaching students problem solving skills have found significant improvements in students' problem solving methods [9,13]. In mathematics, Schoenfeld [57,58] advocates drawing a diagram (if possible) as the first step.

Previous research shows that students who draw diagrams even if they are not rewarded for them are more successful problem solvers [13]. In addition, students who take courses which emphasize effective problem solving heuristics, which include drawing a diagram, are more likely to draw diagrams even on multiple-choice exams [13]. Furthermore, courses which are rich in use of representations can have significant positive impact on student skills [59]. It is therefore possible that explicitly asking students to draw diagrams when solving problems may result in improved performance. An investigation into how spontaneous drawing of free body diagrams (FBDs) affects problem solving [60,61] shows that only drawing correct FBDs improves a student's score and that students who draw incorrect FBDs do not perform better than students who draw no diagrams. Heckler [62] investigated the effects of prompting students to draw FBDs in introductory mechanics by including as the first subpart of each problem an instruction to draw clearly labeled FBDs. He found that students who were prompted to draw FBDs were more likely to follow formally taught problem solving methods rather than intuitive methods, which resulted in deteriorated performance.

Here, we extend the previous research on the impact of drawing diagrams on algebra-based introductory physics (mainly taken by bioscience majors and premeds) student performance on electricity problems through two studies. In study 1, we investigated how student performance was affected when students were provided with a diagram instead of being asked to draw one and compared their performance to that of students who were asked to draw a diagram (without being any more specific than that) and to the performance of a comparison group which was neither asked to draw a diagram nor provided a diagram. We analyzed performance on ten problems throughout the semester given as quizzes; all problems were at the application level of Bloom's taxonomy and they were always related to the topic that was part of the previous week's lecture and the homework that was due (although the quiz problems were not identical to those in the textbook or lecture solved examples or those in the homework). We found that students who were provided diagrams performed significantly worse than the other students on three problems, one which involved standing waves discussed elsewhere [18] and two from electricity discussed here, both of which involve considerations of initial and final conditions. However, in none of the problems analyzed did we find that providing a diagram actually impacted student performance positively. The finding that students provided with a diagram performed worse than students in other groups motivated us to design a follow-up study (study 2), in which we conducted interviews with students with the goal of investigating potential reasons which may account for this finding. Since the first several interviews which were conducted using a think-aloud protocol [63] seemed to impact student reasoning patterns, in a majority of interviews, students were observed by a researcher while they solved the problems posed without being asked to think aloud or disturbed in any other manner. These latter interviews corroborated the findings from the in-class study and suggested one possible reason for the deteriorated performance of students provided with diagrams, namely, that on average, they spent less time on the conceptual analysis and planning stage and jumped into the implementation stage of problem solving without fully understanding the problem.

## II. STUDY 1

### A. Methodology for study 1

A traditionally taught second-semester class of 111 algebra-based introductory physics students was broken up into three different recitations. The three recitations formed the comparison group and two intervention groups for this investigation. All recitations were taught in a traditional manner in which the TA worked out problems similar to the homework problems and then gave a 15 min quiz at the end of class. Students in all recitations attended the same lectures, were assigned the same homework, and took the same exams and quizzes. While the instructor always used effective problem solving strategies, e.g., drawing a diagram, listing knowns or unknowns, making a plan, etc., students were not assessed on whether or not they followed these strategies during quizzes and exams (e.g., no points taken off if students did not draw a diagram). In the recitation quizzes throughout the semester, all students were given the same problems but with the following interventions:

(i) Prompt only group (PO).—in each quiz problem, students were given explicit instructions to draw a diagram with the problem statement.
(ii) Diagram only group (DO).—in each quiz problem, students were provided a diagram drawn by the instructor that was meant to aid in solving the problem.
(iii) No support group (NS).—this group was the comparison group and was not given any diagram or explicit instruction to draw a diagram with the problem statement.

The sizes of the different recitation groups varied from 22 to 55 students because the TA was the same for all three recitations and some students asked the TA if they could go to a different recitation than the one they signed up for. The TA generally allowed students to do this, and the vast majority of students, after choosing the most convenient recitation for their schedule, went to that same recitation every week (e.g., the Tuesday evening recitation). It is also important to note that each intervention was not matched to a particular recitation. For example, in one week, students in the Tuesday evening recitation comprised the comparison group, while during another week the comparison group was a different recitation section. This implies that individual students underwent different interventions from week to week and we therefore do not expect cumulative effects due to the same group of students always being part of the same intervention for the entire semester.

In study 1, we investigated the extent to which asking students to draw a diagram or providing them with one drawn by an expert impacts their problem solving performance. This investigation was carried out for all the quiz problems in a second semester introductory algebra-based physics course. We found that the performance of students provided with a diagram was significantly worse than the performance of students in other groups in two problems from electricity which we discuss below and one problem related to standing sound waves in a tube discussed elsewhere [18].

The two electricity problems are the following (the diagrams provided to students in DO are shown in Figs. 1 and 2):

*Problem 1*
"Two identical point charges are initially fixed to diagonally opposite corners of a square that is 1 m on a side. Each of the two charges $q$ is 3 C. How much work is done by the electric force if one of the charges is moved from its initial position to an empty corner of the square?"

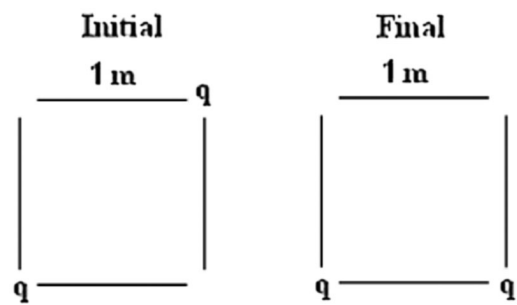

FIG. 1. Diagram for problem 1 provided to students in the DO group.

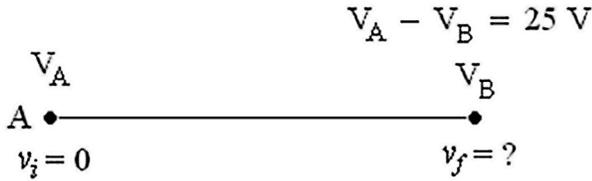

FIG. 2. Diagram for problem 2 provided to students in the DO group.

*Problem 2*

"A particle with mass $10^{-5}$ kg and a positive charge $q$ of 3 C is released from rest from point A in a uniform electric field. When the particle arrives at point B, its electrical potential is 25 V lower than the potential at A. Assuming the only force acting on the particle is the electrostatic force, find the speed of the particle when it arrives at point B."

These diagrams (shown in Figs. 1 and 2) were drawn by the instructor and they are very similar to what most physics experts would initially draw in order to solve the problems (However, the diagrams may be augmented further in the problem solving process as needed). Furthermore, the second diagram also includes an important piece of information from the problem statement that would normally be included in a known quantities or target quantities section of a solution. Neither diagram was meant to trick the students, but rather they were provided as scaffolding support for them. In the interviews conducted in study 2, students were asked to comment on the diagrams and all of them indicated that the diagrams were clear.

In order to ensure homogeneity of scoring, we developed rubrics for each problem analyzed and made sure that there was at least 90% interrater reliability between two independent raters on at least 10% of the data. The development of the rubric for each problem went through an iterative process. During the development of the rubric, the two raters also discussed a student's score separately from the one obtained using the rubric and adjusted the rubric if it was agreed that the version of the rubric was too stringent or too generous. After each adjustment, all students were scored again on the improved rubric. In Table I, we provide the summary of the final version of the rubric used to grade problem 1. The rubric for problem 2 is similar and is included in the Appendix in Table IV.

Problem 1 could be solved by employing two analogous approaches. The first approach (method 1 in Table I) is to use $W = -q\Delta V$ in which $q$ is the charge of the particle and $\Delta V$ is the change in electric potential between the initial and final positions of the charge. The second approach (method 2 in Table I) is to use $W = -\Delta U$ in which $\Delta U$ is the change in the electric potential energy of the configuration of charges between the initial and the final situation. The two approaches are analogous because in both cases one must consider the initial and final situations (charges at opposite corner of the square, charges at adjacent corners of the square) and determine a change in a physical quantity (it is also evident that the two approaches are analogous if one uses the connection between electric potential and electric potential energy, namely, $V = U/q_0$).

Table I shows that for each of the two analogous methods, there are two parts to the rubric: correct and incorrect ideas. Table I also shows that in the correct ideas part, the problem was divided into different sections and points were assigned to each section (10 maximum points). Each student starts out with 10 points and in the Incorrect Ideas part we list the common mistakes students made and how many points were deducted for each of those mistakes. Using the electrostatic force approach to solve this problem is not an effective strategy for students in an algebra-based course (this approach is fairly complex and involves calculus), and students who attempted to use this method did not seem to understand the problem. For example, they would often find the force between the two particles in the initial situation and then multiply this force by 1 m (side of the square). Thus, attempting to use this approach indicated that students had little understanding about how to solve this type of problem, so they were graded separately. The rest of the rubric in the Incorrect Ideas part was used for grading the students who chose a productive approach. For each mistake, we deducted a certain number of points. We note that it is not possible to deduct more points than a section is worth (e.g., the two mistakes that are both labeled 2.1 in Table I are mutually exclusive). We also left ourselves a small window (in the mistake labeled as 2.2) to account for possible mistakes not included explicitly in the rubric.

TABLE I. Summary of the rubric for problem 1 ($W$, $V$, $U$, $q$ refer to work done by the electric force, electric potential, electric potential energy, and charge, respectively).

| | Correct ideas | | |
|---|---|---|---|
| | Method 1 | Method 2 | |
| Section 1 | 1. $W = -q\Delta V$ | 1. $W = -\Delta U$ | 2 p |
| Section 2 | 2. Solve for $V_f$, $V_i$ and find $\Delta V = V_f - V_i$ | 2. Solve for $U_f$, $U_i$ and find $\Delta U = U_f - U_i$ | 7 p |
| Section 3 | 3. Correct units | 3. Correct Units | 1 p |
| | Incorrect ideas | | |
| | Used the electrostatic force incorrectly: if provided correct units (−8 p), if no units (−9 p) | | |
| | Method 1 | Method 2 | |
| Section 1 | 1. Used incorrect equation | 1. Used incorrect equation | −2 p |
| Section 2 | 2.1 Solved for $V_f$ or $V_i$ incorrectly | 2.1 Solved for $U_f$ or $U_i$ incorrectly | −2 p |
| | 2.1 Solved for $V_f$ and $V_i$ incorrectly | 2.1 Solved for $U_f$ and $U_i$ incorrectly | −4 p |
| | 2.2 Did not subtract (−2 p), and/or other mistake (−1 p) | 2.2 Did not subtract (−2 p), and/or other mistake (−1 p) | −3/−1 p |
| | 2.3. Incorrect sign of final answer | 2.3. Incorrect sign of final answer | −1 p |
| Section 3 | 3. Incorrect or no units | 3. Incorrect or no units | −1 p |

### B. Results for study 1

Before discussing the findings for the two problems outlined, we note that the two problems analyzed were part of the same three problem recitation quiz. In the third problem of that quiz, we did not find any statistically significant differences in the performance of the different groups (PO, DO, and NS). Furthermore, students in different groups exhibited almost identical performance on midterm and final examinations.

Table II shows that the average performance of students provided with diagrams was lower by roughly 20% compared to student performance in the other intervention groups. ANOVA [64] indicates that the groups are not comparable ($p < 0.001$) and *post hoc* comparisons between individual groups were conducted to investigate performance differences between groups (we report $p$ values obtained with the Scheffe algorithm for *post hoc* comparisons). We also calculated effect sizes (Cohen's $d$ [64]) for the performance comparisons between different groups. The $p$ values and effect sizes are shown in Table III.

Table III shows that students who were provided diagrams (DO group) performed significantly worse than students in the other two groups. The effect sizes for comparing the performance of students in the DO group with the other groups are quite large, especially for problem 2 where the performance of students in the DO group was on average one standard deviation lower than the performance of students in the other groups. Table III also shows that the performances of students in the PO and NS groups are comparable on both problems. We note that, for problem 1, all students drew a diagram even if they were not specifically asked to do so. However, for problem 2, only 57% of the students in the NS group drew a diagram. But within the NS group, there are no statistical differences between the performance of the students who drew a diagram and those who did not draw a diagram. We performed a $t$ test to compare the performance of students in the NS group who did not draw a diagram and all students in the DO group. We found that students in the DO group performed significantly worse than students in the NS group who did not draw a diagram ($p$ value $= 0.004$, effect size $= 0.788$). Thus, on problem 2, even students who did not draw a diagram performed better than those who were provided a diagram (drawn by the instructor) with the problem statement. Possible reasons for this counterintuitive result were explored and will be discussed in study 2.

TABLE II. Group sizes, averages, and standard deviations for the scores of students in the different groups, out of 10 points.

| Problem 1 | Group size | Average | Standard deviation |
| --- | --- | --- | --- |
| PO (students prompted to draw a diagram) | 26 | 8.5 | 1.9 |
| DO (students provided with a diagram) | 34 | 6.9 | 2.8 |
| NS (students provided with no support) | 51 | 9.0 | 1.4 |
| Problem 2 | Group size | Average | Standard deviation |
| PO (students prompted to draw a diagram) | 26 | 9.0 | 1.4 |
| DO (students provided with a diagram) | 34 | 6.4 | 3.1 |
| NS (students provided with no support) | 51 | 8.6 | 1.3 |

TABLE III. $p$ values (obtained using the Scheffe algorithm) and effect sizes for comparisons between the different groups.

|  | DO-PO | | DO-NS | | PO-NS | |
| --- | --- | --- | --- | --- | --- | --- |
|  | $p$ value | Effect size | $p$ value | Effect size | $p$ value | Effect size |
| Problem 1 | 0.024 | 0.634 | <0.001 | 0.922 | 0.546 | 0.329 |
| Problem 2 | <0.001 | 1.073 | <0.001 | 0.949 | 0.810 | 0.233 |

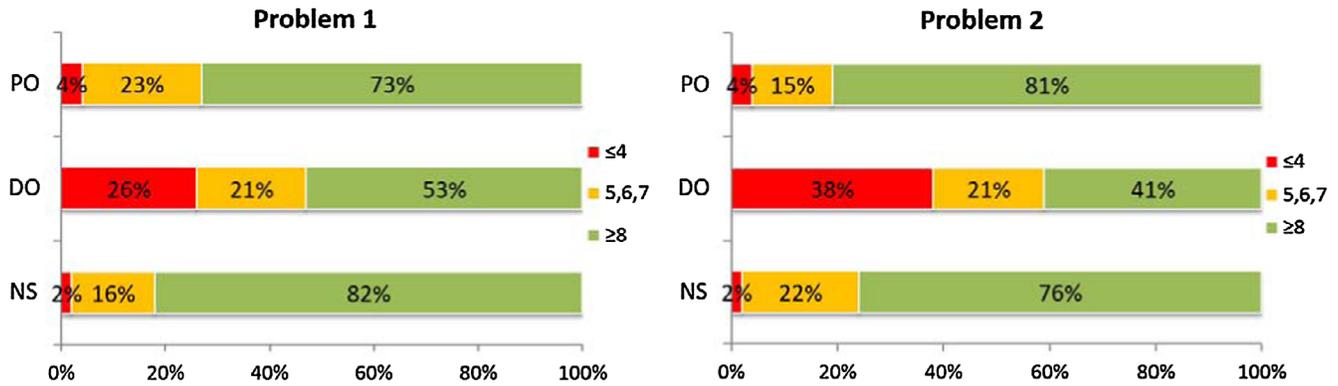

FIG. 3. Percentages of students from each intervention group (PO, DO, and NS) who earned a score of 4 or less, earned a score of 5, 6, or 7, or earned a score above 8.

Furthermore, for each group (PO, DO, and NS), we investigated how many students exhibited poor performance (score of 4 or less out of 10), average performance (score of 5, 6 or 7 out of 10) and good performance (score of 8 or more out of 10). Figure 3 shows the results, which indicate that the percentage of students who performed poorly from the DO group is significantly larger than those in the PO and NS groups on both problems (p values for comparison of the percentage of poorly performing students between DO-PO and DO-NS via Fisher's exact test [64,65] are 0.033 and 0.001, respectively, for problem 1 and 0.002 and <0.001 for problem 2) but the percentages of students with an intermediate score are comparable (all p values identified with Fisher's exact test are larger than 0.5).

## III. STUDY 2

### A. Methodology for study 2

In order to investigate possible reasons for the findings of study 1, interviews were conducted with twenty-three paid student volunteers who were at the time enrolled in an equivalent second semester algebra-based introductory physics course. At the time of the interviews, all students had taken an exam in their course which covered electrostatics and their exam scores varied from below average (e.g., a score of 60 when the class average was 70) to well above average (e.g., a score of 90 when the average is 70). It was not clear *a priori* how the interview protocol would affect students' reasoning and problem solving approaches. For example, it is possible that the think-aloud protocol would alter how students engage in problem solving compared to the case when they are not thinking aloud while solving problems. Therefore, the researchers used one type of interview protocol for some of the students and another type of protocol for another set of students. In particular, six of these interviews were conducted using a think-aloud protocol [63], while in the other seventeen interviews, the students solved the problems while being observed by one of the researchers who took detailed notes of what the students were writing down and at what times (we refer to these latter interviews as the "observational" interviews and the first six interviews as the "think-aloud" interviews). All these interviews took place after students learned and were tested in their course on the relevant concepts required for successfully solving these problems. At the end of all of the interviews, the interviewed students were explicitly asked to comment on the diagrams and how their problem solving processes were impacted by being provided the diagrams in two of the problems.

The goal of the interviews was to investigate the extent to which providing a diagram versus not providing a diagram influences how students engage in problem solving. Thus, in the interviews, students were asked to solve an additional problem, which required the use of the same concepts (conservation of energy or work, electric potential, electric potential energy, etc.) as the two problems discussed in this paper. However, in this additional problem, a diagram was not provided.

*Additional problem*

"A particle of mass $10^{-4}$ kg and charge $q_1 = 1$ $\mu$C is shot at a speed of 10 m/s directly towards another particle with charge $q_2 = 1$ $\mu$C that is held fixed. If the initial distance between the two particles is 1 m, how close does the particle with charge $q_1$ get to $q_2$?"

It is important to note that since problems 1 and 2 both involved considerations of initial and final situations (and were the only two problems from the ones we analyzed which fit this description), the additional problem also involves considerations of initial and final situations. We also made sure that the additional problem was of comparable difficulty to the other problems. We gave the additional problem and problem 2 from study 1 as a quiz to a class of 43 algebra-based introductory students and developed a rubric to score the students' problem solving performance on the additional problem (the rubric was similar to the one shown in Table I and is included in the Appendix in Table V). This rubric was used by four independent raters on 10% of the data and the interrater reliability was better than 90%. For problem 2, the same rubric was used as for study 1. We found that students showed comparable performance on these problems (averages of 7.6 and 7.9 on the additional problem and problem 2, respectively; the $p$ value for comparing students' performance on these two problems was 0.626 and the effect size was 0.105, indicating similar student performance on these two problems). Thus, the additional problem is of comparable difficulty to problem 2, and in study 1, we found that problem 1 is of comparable difficulty to problem 2 (overall averages were 8.04 and 8.27 on problem 1 and problem 2, respectively). Finally, the additional problem was designed such that a student could potentially solve it without drawing a diagram. Out of the 43 students who solved the additional problem, 35% of them did not draw a diagram (and had complete solutions, some correct, some incorrect) indicating that a significant fraction of algebra-based introductory students did not think that drawing a diagram for this problem is necessary in order to solve it.

The goal of the interviews was to compare students' problem solving approaches in the additional problem which did not provide a diagram with the two problems from this study in which the diagrams shown in Figs. 1 and 2 were provided. Since it is unclear if the interview results would be altered if students solve the additional problem first or last, the order in which students were asked to solve the problems was varied: in the first round of interviews (six think-aloud interviews and eleven observational interviews) students solved the additional problem first followed by problems 1 and 2 from study 1, and in the second round of interviews (six observational interviews), students solved problems 1 and 2 from study 1 first followed by the additional problem. In all interviews, students were provided diagrams for problems 1 and 2, but were not provided a diagram for the additional problem. In addition, the interviews were designed to mimic the quiz situation as closely as possible and therefore, students were provided with an equation sheet which was photocopied from the textbook's [66] end of chapter summary (chapter 19, which discusses electrostatic potential and electrostatic potential energy). Students were provided with this equation sheet because in quizzes throughout the semester, the teaching assistant provided students with relevant equations.

### B. Results for study 2

None of the twenty-three interviewed students mentioned anything negative about the diagrams and in general they thought that the diagrams were helpful when provided. A few students noted that they did not necessarily gain anything from being provided the diagrams because if they had not been provided diagrams, they would have drawn something similar. Additionally, the interviews suggested that students were interpreting the diagrams provided in problems 1 and 2 in the intended manner (i.e., they did not find them confusing).

In the six think-aloud interviews, students appeared to approach problems similarly whether or not they were provided a diagram. Three students did not draw a diagram for the additional problem, but they also did not pay much attention to the provided diagrams in problems 1 and 2. The other three students drew diagrams for the additional problem, and in problems 1 and 2, they appeared to pay attention to the diagrams, or drew their own diagram even though one was provided. Also, while reading problems 1 and 2, these students paused to look at the diagram, then read some more, again looked at the diagram, etc. Regarding the approaches to solving the problem, we estimated how much time they spent conceptually analyzing the problem before moving on to the implementation stage. This time was estimated by timing students from when they first started reading the problem statement until they wrote down an equation from the equation sheet provided and started performing algebraic steps. (Note that sometimes students looked at the equation sheet provided and wrote down a formula after which they returned to the equation sheet, or thought about the problem more without performing algebraic steps or writing other formulas down. In cases such as these, the interviewer waited until the student actually started performing algebraic steps to estimate the conceptual planning time.) We found that students spent about the same time conceptually analyzing each of the three problems and also spent about the same time solving each problem. These interviews suggested that partly due to being asked to verbalize their thought process, each student approached the three problems in a very similar manner and was not influenced by being provided diagrams in two of the problems: students often explicitly justified their problem solving approach and tried to provide reasoning. The researchers analyzed the think aloud interviews and concluded that a think-aloud setting did not reproduce quiz conditions very well. In particular, while solving a problem in a quiz, most students do not engage in this type of explicit think aloud reasoning and justification. The researchers realized that asking students to verbalize their thought process in front of a researcher helped motivate students to come up with arguments to support their solutions, and incentivized them to understand the problems they were asked to solve. This conclusion is supported by prior studies as well. For example, in DeVore *et al.*'s study on the challenges in engaging students with self-paced learning tools [67], students were likely to engage much more deeply with a learning tutorial when they were thinking out loud in front of a researcher compared to when working on the tutorial on their own. Students therefore learned significantly more from the tutorials when working while thinking out loud in front of a researcher compared to working on their own. Chi's study on self-explanations [68] found that students who elicited more explanations (remarks related to physics content) when studying worked out examples while thinking aloud showed significantly better performance in solving subsequent problems related to the same content than students who elicited fewer self-explanations. If instead, students are not asked to think aloud while solving a problem or studying worked out examples, they are less likely to self-explain, and some may not do it at all.

While the think-aloud setting did not reproduce the quiz setting, these six think-aloud interviews provided valuable information because they offered further evidence that the additional problem was well chosen. In particular, students spent about the same amount of time conceptually analyzing this additional problem as they did conceptually analyzing the other two problems and they spent about the same amount of time solving this additional problem as they did solving the other two problems. This similarity indicated that the additional problem required a similar amount of time for students to conceptually analyze and complete as the other two problems. Also, the students who had more formula centered approaches to solving problems did not draw a diagram for the additional problem, providing further evidence that some students did not consider that drawing a diagram was necessary or helpful to solve the additional problem.

In the first eleven observational interviews, students solved the additional problem (which did not provide a diagram) followed by problems 1 and 2 (which provided diagrams) from study 1 and in the last six observational interviews conducted, the order of the problems was switched and students first solved problems 1 and 2 from study 1 (diagrams provided) and then solved the additional problem (diagram not provided).

It is important to stress that this design ensures that each student acts as their own control. In other words, the determination of whether a student spends more or less time conceptually analyzing a problem which has a diagram provided compared to another problem which does not provide a diagram was done for each student. This is important because different students may have different problem solving strategies and may end up spending different amounts of time conceptually analyzing a problem. But if each student acts as their own control, we can draw conclusions about the extent to which a provided diagram may influence the amount of time a student spends conceptually analyzing a problem.

During these seventeen observational interviews, roughly half the students (nine students) started the implementation stage of the problem solving process while solving problems 1 and 2, in which the diagrams were provided, noticeably earlier than when solving the additional problem in which a diagram was not provided. This was found both for students who solved the additional problem first (six out of eleven interviewed students) and for students who solved the additional problem last (three out of six interviewed students). Including all seventeen students, we found that on average, students spent 71% more time conceptually planning the additional problem (which did not provide a diagram) than the other two problems. In a few of those cases, in one problem or the other, this quicker focus on manipulation of equations appeared to negatively impact their performance.

Suzana (an interviewed student) had received an 83% on the electricity exam, and an A in the first semester physics class. She solved the problems in the order (i) additional problem (no diagram provided), (ii) problem 1 (diagram provided), (iii) problem 2 (diagram provided). Her work for the additional problem and problem 2 are shown in Fig. 4 (in her work, EPE refers to electric potential energy, or $U$). While solving the additional problem (given first) in which a diagram was not provided, it appeared that Suzana was aware that electric potential and electric potential energy are different because she used the equation which relates these two quantities, $EPE = q_0 V$ and explicitly solved for the electric potential energy of two charges to obtain $EPE = kqq_0/r$. She also correctly used electric potential energy to solve the additional problem and it was apparent from the interview that she used the resources for electric potential energy and electric potential appropriately. On the other hand, while solving problem 2, Suzana immediately wrote down two electric potential energies: $EPE_A = 25$ V and $EPE_B = 0$ V, even though the diagram provided contained an equation relating electric potentials ($V_A - V_A = 25$ V), not electric potential energies. And despite the fact that she used the resources of electric potential and electric potential energy appropriately in a previous problem (which did not provide a diagram), she appeared to have difficulty distinguishing between them in this problem (which did provide a diagram).

Another student, Calvin, solved the problems in the order (i) problem 1 (diagram provided), (ii) problem 2 (diagram provided), (iii) additional problem (no diagram provided). In the first two problems, Calvin treated the electric potential as electric potential energy. For example, in problem 2 (work shown in Fig. 5), after a false start with electric potential, he attempted to use conservation of energy and wrote $KE_{final} = 25$ V even though 25 V is given to be the change in electric potential (the diagram also explicitly shows the information that $V_A - V_B = 25$ V). It appeared that Calvin was attempting to use information provided in the problem before analyzing the problem qualitatively and ensuring he understood what the information really meant. In the additional problem, on the other hand (work shown in Fig. 6), Calvin spent significantly more time and, while he also used conservation of energy in that problem, he used the correct expression for the electric potential energy of two charges separated by a distance $r$: $U = kq_1q_2/r$.

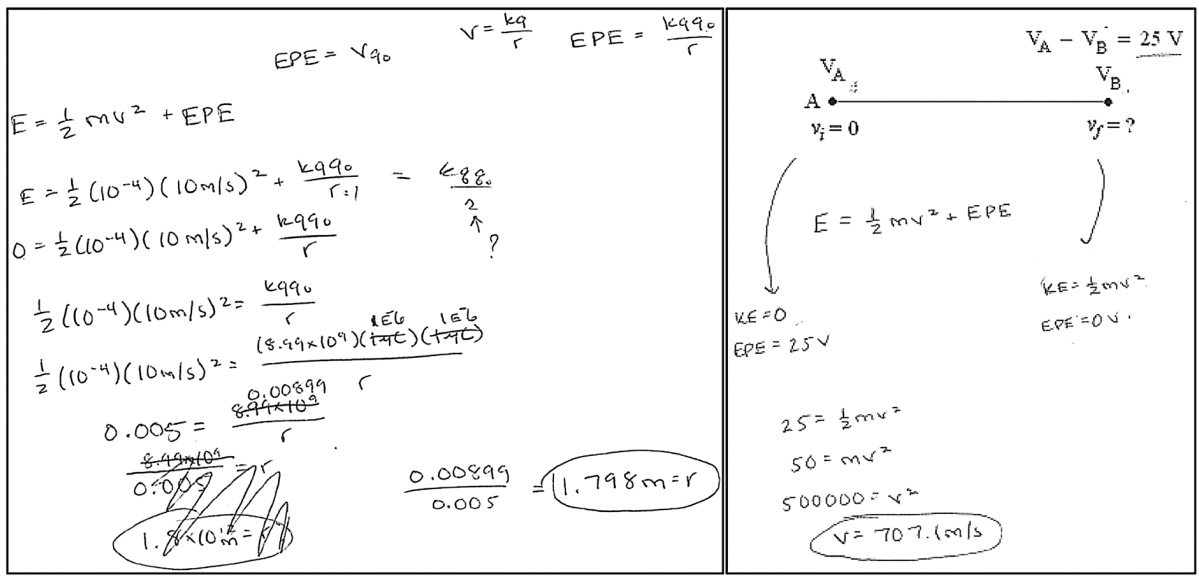

FIG. 4. Suzana's solution to the additional problem (left) on which she worked first and problem 2 (right), on which she worked last.

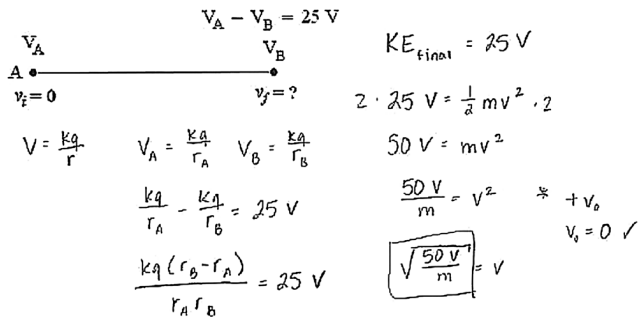

FIG. 5. Calvin's work to problem 2 (second problem solved in interview).

Interviews suggested that these students proceeded to manipulate equations earlier in the problems which provided a diagram and they did not spend sufficient time conceptually analyzing these problems. In Suzana's case, despite the fact that she had previously realized while solving the first problem that electric potential energy and electric potential are different, in problem 2 she did not appear to distinguish between them which resulted in an incorrect solution. In Calvin's case, he also realized that the electric potential and electric potential energy are different only while solving the problem which did not provide a diagram.

In fact, similar to these two students, seven other interviewed students, almost immediately after reading this problem, which included a diagram, started looking at the equation sheet. Then, they often copied a formula on their paper and proceeded to solve the problems using the formula, which sometimes negatively impacted their performance on these problems.

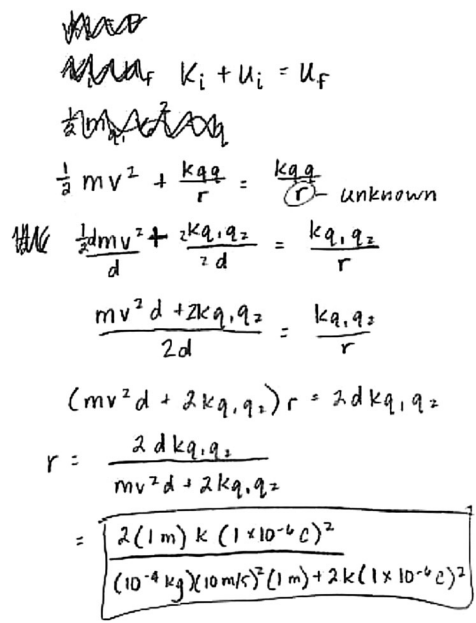

FIG. 6. Calvin's work to the additional problem (last problem solved in interview).

The observational interviews suggest that roughly half of the students were spending less time conceptually analyzing the problems in which diagrams were provided. These students appeared to jump to the implementation stage of the problem solving process immediately, which sometimes had a negative impact on their performance.

## IV. DISCUSSION AND SUMMARY

Prior research suggests that students in classes which promote conceptual understanding through active-learning methods outperform those from traditionally taught classes even on quantitative tests [25,50,69–75]. This finding suggests that students who perform poorly on physics problem solving may do so not because they have poor mathematical skills, but rather because they do not effectively analyze the problem conceptually. In particular, they may not employ effective problem solving heuristics and transform the problem into a representation which makes further decision making and consideration of relevant physics principles easier. For example, converting a physics problem from the verbal to the diagrammatic representation by drawing a diagram is a heuristic that can facilitate better understanding of the problem and aid in solving it.

In study 1, we found that students who were provided diagrams performed significantly worse on two electrostatics problems than students who were not, and the quantitative data suggested that a much larger percentage of students provided with diagrams did not understand the problem. This in turn suggests that they had difficulty conceptualizing the problem and formulating a correct solution. The fact that many students who were provided diagrams failed to understand the problem conceptually (leading to very poor performance) was also evident from observing their individual solution strategies. For example, students provided with diagrams were more likely than students in the other groups to employ formula-based approaches and it was sometimes unclear by observing their written work how they arrived at the decision to use those formulas (which were sometimes not productive for the problems).

In study 2, we conducted interviews with students who solved the two problems (which provided diagrams) as well as an additional problem on the same topics (which did not provide a diagram) while being observed by a researcher. The interviews suggested that students who are provided with a diagram may spend less time conceptually analyzing the problem and jump into the implementation stage before understanding the problem. In the interviews, roughly half the students (nine out of seventeen) spent considerably less time thinking about the problem conceptually when a diagram was provided compared to when it was not. Some of these students looked at the equation sheet almost immediately after reading the problems which provided a diagram, but in the problem which did not provide a diagram, they spent more time thinking about the problem first (performing some sort of conceptual analysis or trying to understand the physical situation presented) before looking for a relevant equation to use.

We should point out that none of the ten problems given in quizzes with or without diagrams throughout the semester had physical situations which were very complex or required long descriptions, and introductory students were expected to be able to picture these physical situations on their own without support. Also, in electrostatics, the two problems in which we found deteriorated performance of students provided with a diagram were the only ones which involved considerations of initial and final situations. It is possible that these types of problems may be more susceptible to a negative effect due to providing diagrams. Also, we cannot say anything conclusive about why student performance was not affected by providing diagrams in the other problems since our investigation discussed here in study 2 focused on the two problems in which providing a diagram resulted in deteriorated performance. However, in none of the 10 quiz problems did providing a diagram result in improved student performance. Thus, our study suggests that for problems that are not very complex, providing a diagram (i) may have a detrimental effect, and (ii) is unlikely to have a beneficial effect. The findings of this study suggest that instructors should avoid providing diagrams to introductory physics students for problems which students can reasonably be expected to understand from a verbal description alone and draw a diagram themselves. This is because the process of translating a problem from a verbal representation into a diagrammatic one is important for the initial conceptual planning stage of problem solving—something many introductory students skip without explicit guidance and support, and when they do skip this important stage, it can lead to deteriorated performance.

It is also important to note that students who were asked to draw diagrams were almost always statistically more likely to draw productive diagrams (as defined from an expert's point of view) than students in the other intervention groups. Within a cognitive apprenticeship model [76], asking students to draw diagrams is a type of scaffolding support, and this investigation indicates that students asked to draw a diagram did not perform worse than those provided with no support on typical quiz problems (i.e., problems that are not very difficult or complex). Therefore, an implication of the study reported here based on the cognitive apprenticeship model is that in order to help students learn the usefulness of drawing diagrams in problem solving, students can be asked to draw diagrams in various assignments and quizzes throughout the semester. This support can be reduced as students begin drawing more diagrams and recognize their usefulness on their own. Moreover, since assessment drives learning [77], it is likely that rewarding students for drawing appropriate diagrams will have a beneficial effect. This can be one helpful step in getting students accustomed to using productive problem solving heuristics, and over time making them better at performing initial conceptual analysis and planning of the problem solution on their own.

Finally, prompted by the results of study 1, we had discussions with ten instructors who regularly teach introductory physics and they nearly always conjectured that providing students with diagrams would likely lead to improved performance. Our study suggests that providing a diagram was never helpful for students, and, in certain cases as discussed here, it can actually be detrimental. This discrepancy between instructor predictions and student performance suggests that the manner in which providing diagrams for these two problems, which involve considerations of initial and final conditions affects students' performance, is quite complex and not at all intuitive.

## ACKNOWLEDGMENTS

We would like to thank the National Science Foundation for Grant No. 1524575 and the members of the physics education research group at the University of Pittsburgh (E. Marshman, S. DeVore) as well as R. P. Devaty for useful discussions and feedback on the manuscript.

## APPENDIX: WORKED OUT SOLUTIONS TO PROBLEMS 1 AND 2 AND RUBRICS USED FOR PROBLEM 2 AND THE ADDITIONAL PROBLEM

Figure 7 shows an instructor worked out solutions to problems 1 and 2 as well as how the points were allocated to various parts of the solutions.

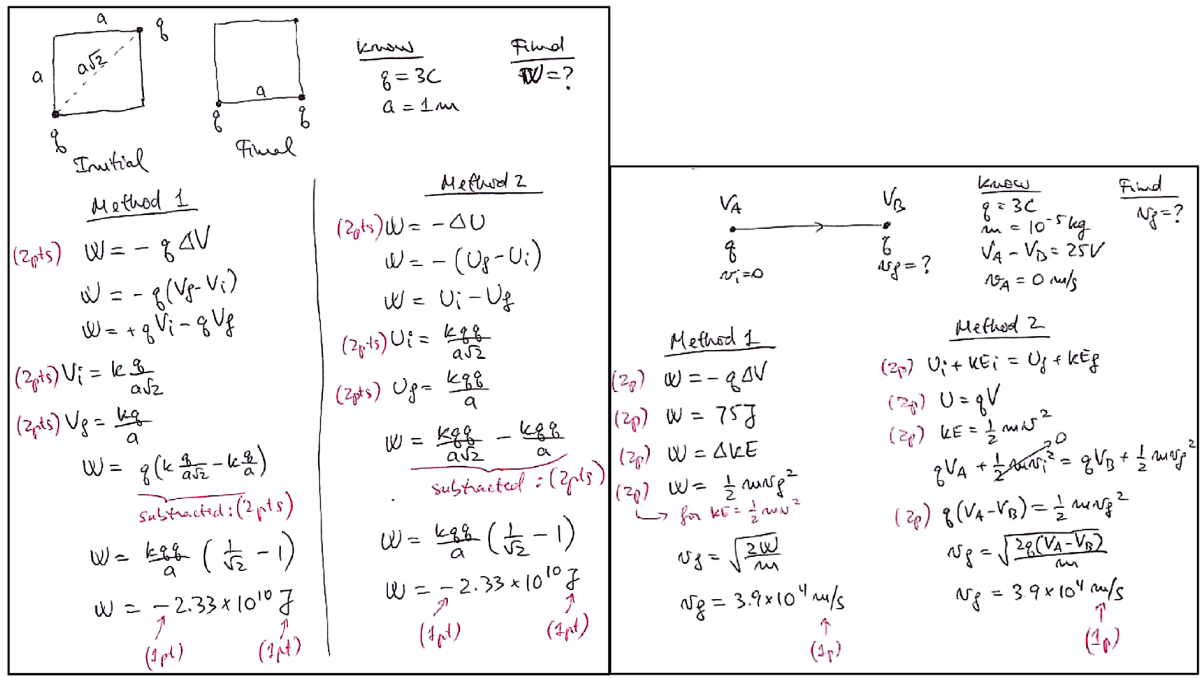

FIG. 7. Worked out solutions for problems 1 (left) and 2 (right) along with how many points were assigned to each part of the solution. Both problems can be solved using two equivalent methods and both methods are shown.

TABLE IV. Rubric used to score students' problem solving performance on problem 2.

| | Correct ideas | | | Correct ideas | |
|---|---|---|---|---|---|
| | Method 1: work-energy theorem | | | Method 2: conservation of energy | |
| Section 1 | $W = -q\Delta V$ | 2 p | Section 1 | $U_i + KE_i = U_f + KE_f$ | 2 p |
| | Calculated $W$ correctly | 2 p | | $qV_A - qV_B = \Delta KE$ | 2 p |
| Section 2 | $W = \Delta KE$ | 2 p | | $U = qV$ | 2 p |
| | $KE = \frac{1}{2}mv^2$ | 2 p | | $KE = \frac{1}{2}mv^2$ | 2 p |
| Section 3 | Correct units in final answer | 1 p | Section 2 | Correct units in final answer | 1 p |
| | Incorrect ideas | | | Incorrect ideas | |
| Section 1 | Used incorrect equation | −2 p | Section 1 | Used incorrect equation | −2 p |
| | Calculated $W$ incorrectly[a] | −2 p | | Calculated $\Delta KE$ incorrectly | −2 p |
| | Obtained the wrong sign for $W$ | −1 p | | Obtained the wrong sign for $\Delta KE$ | −1 p |
| Section 2 | Used incorrect equation connecting the two parts | −2 p | | Did not use $U = qV$, or used incorrect equation | −2 p |
| | Used incorrect equation for $KE$ | −2 p | | Used incorrect equation for $KE$ | −2 p |
| Section 3 | Incorrect or no units | −1 p | Section 3 | Incorrect or no units | −1 p |

[a]If a student uses an incorrect equation, but uses it correctly, these two points are not taken off.

TABLE V. Rubric used to score students' problem solving performance on the additional problem (used in study 2).

| | Correct ideas | |
|---|---|---|
| Section 1 | Conservation of energy | 3 p |
| Section 2 | Calculating $E_i$, $E_f$, and solving for $r_f$ | 6 p |
| Section 3 | Units + sanity check[a] | 2 p |
| | Incorrect ideas | |
| Section 1 | Did not use conservation of energy | −3 p |
| Section 2 | Expression for initial energy is missing both $KE$ and $U$ | −4 p |
| | Expression for initial energy has only $KE$ (but correct) | −3 p |
| | Expression for initial energy has only $U$ (but correct) | −2 p |
| | Expression for initial energy has both $KE$ and $U$, but one incorrect | −1 p |
| | Expression for final energy is incorrect | −2 p |
| Section 3 | No or incorrect units | −1 p |
| | Answer for $r_f$ that does not make sense (with no comment[b]) | −1 p |

[a]A sanity check was important here because many students found $r_f > 1$ m, which does not make sense.
[b]If a student obtains an answer that does not make sense, but make a statement about it, he/she is awarded the point for the sanity check.